\documentclass[intlimits,twoside,a4paper]{article}
\usepackage{amsmath,amssymb,amsfonts,epsfig,graphicx,colordvi}
\usepackage{colordvi}
\usepackage{amsmath,amssymb}
\usepackage{graphicx}

\usepackage[T2A]{fontenc}
\usepackage[cp1251]{inputenc}

\usepackage[eqsecnum]{cmpj2}
\newcommand {\lp}{{\rm 1\hskip-0.09cm l}}

\issue{2015}{18}{3}{33003}
\doinumber{10.5488/CMP.18.33003}

\title[The $\beta$-expansion of the $D=1$ fermionic  spinless Hubbard model off the half-filling regime]%
{The $\beta$-expansion of the $D=1$ fermionic  spinless Hubbard model off the half-filling regime}

\author[E.V. Corr\^ea Silva, M.T. Thomaz, O. Rojas]%
{E.V. Corr\^ea Silva\refaddr{lb1}, M.T. Thomaz\refaddr{lb2}\thanks{E-mail: mtt@if.uff.br}\,, O. Rojas\refaddr{lb3}}
\addresses{
\addr{lb1} Departamento de Matem\'atica, F\'{\i}sica e Computa\c{c}\~ao,  Faculdade de Tecnologia, Universidade do Estado do Rio de Janeiro,   Rodovia Presidente Dutra km 298 s/n$^{\textit o}$, P\'olo  Industrial, CEP 27537--000,  Resende-RJ, Brazil
\addr{lb2} Instituto de F\'{\i}sica, Universidade Federal Fluminense, Av. Gal. Milton Tavares de Souza s/n$^{\textit o}$,  CEP 24210-346, Nite\'oi-RJ, Brazil
\addr{lb3} Departamento de F\'{\i}sica, Universidade Federal de Lavras, Caixa Postal 3037, CEP 37200-000, Lavras-MG,  Brazil
}

\authorcopyright{E.V. Corr\^ea Silva, M.T. Thomaz, O. Rojas, 2015}
\date{Received May 15, 2015}

\begin{document}

\maketitle

\begin{abstract}
We found that when  the spinless model is
off the half-filling regime ($\mu \neq V$),
the Helmholtz free energy (HFE)
can be written  as two $\beta$-expansions:
one expansion comes from the half-filling configuration
and another one that  depends on the parameter
$x = \mu - V$. We show numerically that the
chemical potential as a function of temperature satisfies a relation
similar to the one derived  from the particle-hole symmetry of
the fermionic spinless  model. We extend the $\beta$-expansion
of the HFE of the one-dimensional  fermionic spinless
Hubbard model up to order $\beta^8$.
\pacs 05.30.Fk, 71.27.+a, 75.10.Pq
\end{abstract}



\section{Introduction} \label{S1}

One-dimensional models are certainly
easier to handle than higher-dimensional ones,
and for a long time they have been treated as
toy models. In general, these models are a simplified description of a
real physical system.  It is often difficult to realize
what is missing in those simple models in order to explain
the experimental results.

The development of optical lattices over the last two decades has made possible the physical realization of one-dimensional models like the
spin-$1/2$ Ising model \cite{simon}, thus offering
the opportunity  for the experimental verification of the
predictions of simplified models like the one-band Hubbard
model \cite{hubbard1,hubbard2}, that
partially describes quantum magnetic phenomena.

The simplest one-dimensional fermionic model is the fermionic
spinless Hubbard model,  the generalizations  of which have been
applied to  the description of Verwey metal-insulator transitions
and charge-ordering phenomena
of Fe$_3$O$_4$, Ti$_4$O$_7$,  LiV$_2$O$_4$ and
other $d$-metal compounds \cite{verwey,kobayashi,fulde}.

In references \cite{haldane,sznajd} it is shown that the
fermionic spinless  Hubbard model in $D=1$ is mapped
onto the exactly soluble  $D=1$ spin-$1/2$ $XXZ$ Heisenberg
model in the presence of a longitudinal magnetic field.
The fermionic model has a particle-hole symmetry \cite{sznajd}.
In reference \cite{comment} we explore the consequences of
that symmetry on the thermodynamic functions of this model in the
whole interval of temperature $T>0$.

The spin-$1/2$ $XXZ$ Heisenberg model is an exactly
solvable model. Its thermodynamics can be derived from
the thermodynamic Bethe ansatz equations \cite{livro_takahashi}.

B\"uhler {et al.} calculated the $\beta$-expansion
of the specific heat and the susceptibility, both per site,
of the frustrated and unfrustrated  spin-$1/2$ Heisenberg
chain up to order $\beta^{16}$  and $\beta^{24}$, respectively,
in the absence of an external magnetic field \cite{buhler}
[$h=0$ on the r.h.s. of  equation (\ref{2})].
In 2001 Takahashi derived an integral  equation to obtain the HFE
of the spin-$1/2$ $XXZ$ model \cite{takahashi2001}.
The high temperature expansions of the specific heat  and
the susceptibility, both per site, of the isotropic spin-$1/2$
$XXX$ model \cite{shiroishi} were calculated up to order
$\beta^{100}$ also  for $h=0$. In the language of the
 spinless model, the  absence of a magnetic
field in the spin-$1/2$ model corresponds to the half-filling case.

In reference \cite{EPJB2005} we calculated the  $\beta$-expansion
of the  Helmholtz free energy (HFE) of  the one-dimen\-sional
spin-$S$ $XXZ$ Heisenberg model in the presence of a
longitudinal magnetic field,
$S \in \{ \frac{1}{2}, 1, \frac{3}{2}, \dots$\}
up to order $\beta^6$. By applying  the mapping  between
the aformentioned one-dimensional fermionic  and spin models,
we obtain the expansion of the HFE of the fermionic spinless
Hubbard model also up to order $\beta^6$.  These high
temperature expansions are analytic and valid for any set
of parameters of the  respective Hamiltonian, thus letting one
avoid the numerical solution of a hierarchy of coupled  integral for
every set of parameters of the spin-$1/2$ $XXZ$ model.

In the present article
we study the $\beta$-expansion of
thermodynamic  functions of the spinless Hubbard model off the
half-filling regime.  We calculate two
additional orders in the $\beta$-expansion of the HFE of
reference \cite{EPJB2005}  and verify the consequences of
those extra terms on the specific heat per site and
on the mean number of spinless fermions per site.  We also
numerically study the dependence  of the chemical potential on
the temperature when the number of particles in the chain
is fixed.

In section~\ref{S2} we present the  Hamiltonian of the
one-dimensional fermionic  spinless Hubbard model and its mapping
onto the $D=1$ spin-$1/2$ $XXZ$ Heisenberg model in the
presence of a longitudinal magnetic field. We
present the relations satisfied  by the HFE of the model
due to the particle-hole symmetry. In section~\ref{S3}
we discuss  the $\beta$-expansion of the specific heat
per site and the mean number of spinless fermions per site
off the half-filling regime, and show the parameters of
expansions of thermodynamic functions.
In section~\ref{S4}
we use the $\beta$-expansion of the mean number of
spinless fermions per site to numerically discuss
the dependence of the chemical
potential on temperature when the number of fermions
in the chain is kept constant.  Finally, section~\ref{S6} has a summary of our results. Appendix~\ref{Apend_A}
has the $\beta$-expansion of the HFE of the fermionic spinless
Hubbard model in $D=1$, up to order $\beta^8$.


\section{The  fermionic spinless Hubbard model in $D=1$
and its exact relations }  \label{S2}

The fermionic spinless Hubbard model in $D=1$ is a
very simple anti-commutative model whose Hamiltonian is \cite{sznajd}:
\begin{subequations}
\begin{eqnarray}
{\bf H} (t, V, \mu) = \sum_{i=1}^{N} \; {\bf H}_{i, i+1}  (t, V, \mu) , \label{1a}
\end{eqnarray}
\noindent in which
\begin{eqnarray}
{\bf H}_{i, i+1}  (t, V, \mu) \equiv  t  ({\bf c}_i^{\dag} {\bf c}_{i+1}
  +  {\bf c}_{i+ 1}^{\dag} {\bf c}_{i}) + V  {\bf n}_i {\bf n}_{i+1}
      - \mu  {\bf n}_i  \,,   \label{1b}
\end{eqnarray}
\end{subequations}
the operators ${\bf c}_i $ and ${\bf c}_i^{\dag}$,  with
$i\in\{ 1, 2, \dots, N\}$, are the destruction and creation fermionic operators,
respectively,  and $N$ is the number of
sites in the periodic chain (${\bf H}_{N, N+1} =  {\bf H}_{N,1}$).
Those operators satisfy anti-commutation relations,
$ \{ {\bf c}_i, {\bf c}_j^{\dag} \} = \delta_{i j} \lp_i$
and $\{{\bf c}_i, {\bf c}_j \} = 0$. In this Hamiltonian  $t$ is
the hopping integral, $V$  is the strength of the repulsion ($V>0$)
or attraction ($V<0$) between first-neighbour fermions, and $\mu$ is the
chemical potential. The operator number of fermions at the $i^{\rm th}$
site of the  chain is defined as
${\bf n}_i \equiv {\bf c}_i^{\dag}  {\bf c}_i$.

It is shown in the literature \cite{haldane,sznajd}  that the
equivalence of the Hamiltonian (\ref{1a})--(\ref{1b}) and the one that
describes the spin-$1/2$ $XXZ$ Heisenberg model in $D=1$,
\begin{eqnarray}  \label{2}
{\bf H}_{S = 1/2}(J, \Delta, h) =  \sum_{i=1}^{N} \left[ J \left( {\bf S}_i^x {\bf S}_{i+1}^x
+ {\bf S}_i^y {\bf S}_{i+1}^y  + \Delta {\bf S}_i^z {\bf S}_{i+1}^z  \right)
      - h {\bf S}_i^z\right]  ,
\end{eqnarray}
in which ${\bf S}^l = {\sigma^l}/{2}$,
$l\in\{ x, y, z\}$, and $\sigma^l$  are the Pauli
matrices;   the parameters of both Hamiltonians  satisfy
the relations:
\begin{eqnarray}  \label{3}
J = 2t  ,  \qquad   \Delta = \frac{V}{2t}
\qquad \text{and} \qquad
 h = \mu - V  .
\end{eqnarray}

The Hamiltonians (\ref{1a})--(\ref{1b}) and (\ref{2}), with their
parameters satisfying conditions (\ref{3}),  differ by a
constant operator
\begin{eqnarray}  \label{4}
 {\bf H} (t, V, \mu) = {\bf H}_{S = 1/2}(J = 2t, \Delta ={V}/{2t}, h = \mu - V)
     - N \left( \frac{J \Delta}{4} + \frac{h}{2} \right)  \lp,
\end{eqnarray}
in which $\lp$ is the identity operator of the chain.

Let ${\cal Z} (t, V, \mu; \beta)$  and ${\cal Z}_{S = 1/2} (J, \Delta, h ; \beta)$
be the partition functions of the fermionic spinless  model and the spin chain model,
respectively,
\begin{subequations}
\begin{eqnarray}
{\cal Z} (t, V, \mu; \beta) &=&  \textrm{Tr}\left\{ \re^{-\beta {\bf H} (t, V, \mu)}  \right\} ,
          \label{5a}     \\
{\cal Z}_{S = 1/2} (J, \Delta, h ; \beta)
       &=&  \textrm{Tr}\left\{ \re^{-\beta {\bf H}_{S = 1/2}(J, \Delta, h)}  \right\}  , \label{5b}
\end{eqnarray}
\end{subequations}
in which $ \beta = {1}/{kT}$,
$k$ is the Boltzmann's constant
and $T$ is the absolute temperature in kelvin.

The  functions  $W (t, V, \mu; \beta)$  and
$W_{S = 1/2} (J, \Delta, h ; \beta)$ are
the HFE's associated to the Hamiltonians (\ref{1a})--(\ref{1b})
and (\ref{2}), respectively, in the thermodynamic limit
($N \rightarrow \infty$)
\begin{subequations}
\begin{eqnarray}
W (t, V, \mu; \beta) &=& -  \lim_{N \rightarrow \infty}  \frac{1}{N} \frac{1}{\beta}
           \ln \left[{\cal Z} (t, V, \mu; \beta) \right]   ,   \label{6a} \\
W_{S = 1/2} (J, \Delta, h ; \beta) &=& -  \lim_{N \rightarrow \infty}
 \frac{1}{N} \frac{1}{\beta}
     \ln \left[{\cal Z}_{S = 1/2} (J, \Delta, h ; \beta) \right]   ,   \label{6b}
\end{eqnarray}
\end{subequations}
in which $N$ is the number of sites in the chain.

Due to the equality of operators in  equation (\ref{4}),  we have
a relation between the HFE's (\ref{6a}) and (\ref{6b}) \cite{sznajd},
\begin{eqnarray}   \label{7}
W (t, V, \mu; \beta) = W_{S = 1/2}(J = 2t, \Delta = {V}/{2t}, h = \mu - V; \beta)
     + \left( \frac{V}{4} - \frac{\mu}{2}  \right)  ,
\end{eqnarray}
valid at any non-null temperature $T$. This relation
permits to relate the thermodynamic functions of both one-dimensional
models.

The expression of the function $W_{S = 1/2} (J, \Delta, h ; \beta)$
comes from the calculation of the trace of the operator
$\re^{-\beta {\bf H}_{S = 1/2}(J, \Delta, h)}$ over all sites in the chain.
In the $\beta$-expansion of this function, only  terms with an
even number of operators ${\bf S}_i^z$ at
each site  give a non-null value to the trace at the $i^{th}$ site,
and, therefore, we obtain that  the HFE of the one-dimensional
$S=1/2$ $XXZ$ Heisenberg model is an even function of the longitudinal
magnetic field $h$,
\begin{eqnarray}  \label{8}
  W_{S = 1/2}(J, \Delta, - h; T) =  W_{S = 1/2}(J, \Delta,  h; T) .
\end{eqnarray}
Another way to understand the invariance (\ref{8})
of  $ W_{S = 1/2}$ is to remember the  symmetry
of the Hamiltonian (\ref{2})  upon reversal of the external magnetic field,
$h \rightarrow -h$, and of the spin operators,
$\vec{\bf S}_i \rightarrow -\vec{\bf S}_i$, in which
$i \in \{1, 2, \dots N\}$.

Consider, for a given magnetic field $h$ and a fixed value (positive,
null or negative) of $V$, the chemical potential $\mu$ so that $ h = \mu- V$.
For a reversed magnetic field, the corresponding chemical potential
$\mu_2$ for which $-h = \mu_2 - V$ is
\begin{eqnarray}  \label{9}
   \mu_2 = - \mu + 2V  .
\end{eqnarray}

The identity (\ref{8}) and the condition (\ref{9})
recover  the symmetry particle-hole  of the fermionic
spinless Hubbard model for any values of $V$ and $\mu$.
This symmetry is summarized in the relation of the HFE of
the fermionic spinless model at the same potential $V$ and
different chemical potentials,
\begin{eqnarray}  \label{10}
  W (t, V, \mu; \beta) = W (t, V, \mu_2 = -\mu + 2V; \beta) - (\mu -V)  .
\end{eqnarray}

In reference \cite{comment} we explore the effect  of the relation
(\ref{10}) on the thermodynamic functions  of the one-dimensional
fermionic spinless model at the same potential $V$ but with
chemical potentials $\mu$ and $\mu_2$. The results discussed
in reference \cite{comment} are valid in the whole range of temperatures
of $T>0$.

In reference \cite{EPJB2005} we use the method of reference \cite{chain_m}
to calculate the $\beta$-expansion of the spin-$S$ $XXZ$ Heisenberg
model in $D=1$, in the presence of a longitudinal magnetic field up
to order $\beta^6$, with $S \in\{ \frac{1}{2}, 1, \frac{3}{2}, \dots\}$
For a summary of the results of reference \cite{chain_m} we suggest to
the reader  reference \cite{winder}. Relation (\ref{7})
permits to derive the HFE of the chain of  spinless
fermions from the $\beta$-expansion presented
in reference \cite{EPJB2005} up to order $\beta^6$.

In this article we introduce  a new set of rules  for  algebraic
calculation using  the method of reference \cite{chain_m} that enables
us to calculate the $\beta$-expansion of the HFE of the fermionic
spinless  Hubbard model in $D=1$  up to order $\beta^8$.

In equation (\ref{A.1}) we present the $\beta$-expansion of the HFE of the
one-dimensional fermionic spinless Hubbard model up to order $\beta^8$.
This result is calculated using the method of reference \cite{chain_m} for
arbitrary values of the parameters
in the Hamiltonian (\ref{1a})--(\ref{1b}).
The coefficient of the $\beta^n$ term, with
$n \in \{ -1, 0, 1, \dots, 8\}$,
in expansion (\ref{A.1}) is exact.
The polynomial form of the HFE expansion in
$\beta$ and in the parameters of the Hamiltonian (\ref{1b})
can be easily handled by any computer algebra system.
Thermodynamic functions of the model can be derived
from the appropriate derivatives of the  HFE.

We explicitly verified that expansion  (\ref{A.1}) satisfies
the relation (\ref{10}), which is valid  separately for each
coefficient of the $\beta^l$ terms of this HFE, with
$l \in \{1, 2,\dots, 8\}$.


\section{Discussion on the $\beta$-expansion of the HFE of the
          model}   \label{S3}

The $\beta$-expansion (\ref{A.1}) of the HFE of the fermionic spinless
Hubbard model in $D=1$ permits the derivation of
the $\beta$-expansion of various thermodynamic
functions. In this article we discuss only two thermodynamic
functions: the specific heat per site $ {\cal C} (t, V, \mu; \beta)
=  - \beta^2  {\partial^2 [ \beta W]}/{\partial \beta^2}$,
and the mean number of spinless fermions per site
$\rho ( t, V, \mu; \beta) = - {\partial W}/{\partial \mu}$.
(From this point on, it will be ommitted that those functions
are calculated per site.) The expansion (\ref{A.1}) is two orders
higher in $\beta$ than the $\beta$-expansion of the HFE of the
one-dimensional spin-$1/2$ $XXZ$  Heisenberg model, in the
presence of a longitudinal magnetic field, presented in reference \cite{EPJB2005}.
In what follows we make a simple comparison,
the $\beta$-expansions of the specific heat
and the mean number of particles, derived from the expansion
of the HFE in reference \cite{EPJB2005} and equation (\ref{A.1}),
are compared to their respective exact expressions
of two simple limiting cases, and the interval of $\beta$
in which there is a good agreement between them is
determined.

In order to verify the range of convergence of each expansion, we
compare them to the respective thermodynamic function of two
limiting cases of the Hamiltonians (\ref{1a})--(\ref{1b}) and (\ref{2}):
the free spinless fermion model \cite{BJP2001}  and the spin-$1/2$
Ising model in the presence of a longitudinal magnetic
field \cite{baxter}. We do not need any extra computational
effort to exactly calculate these two limiting cases for
arbitrary values of the parameters in their respective Hamiltonians.

Let ${\cal C}_7$ and ${\cal C}_9$ be the specific heat and the
$\beta$-expansion up to order $\beta^7$ and $\beta^9$,
respectively, derived from the HFE of reference \cite{EPJB2005}
and equation (\ref{A.1}). We have compared the expansions
${\cal C}_7$ and ${\cal C}_9$ to the  specific heat of the
free spinless fermion model \cite{EPJB2005}
and the spin-$1/2$ Ising model \cite{BJP2001,baxter}, both in $D=1$.
In order to measure the difference between each  specific heat
of the exactly soluble models and its expansions ${\cal C}_7$ and
${\cal C}_9$,  we define the percentage difference,
\begin{eqnarray}  \label{14}
\delta_D {\cal C}_k \equiv 100\% \times
   \left( \frac{{\cal C}_M - {\cal C}_k}{{\cal C}_M} \right)  ,
      \hspace{1cm}
	         k \in \{7,\ 9\},
\end{eqnarray}
with  $M \in \{ \text{Ising}, \ \text{Free}\}$.
Let ${\cal C}_\textrm{Ising}$  and ${\cal C}_\textrm{Free}$ be the specific heat
of the spin-$1/2$ Ising model and that of the free
spinless fermion model, respectively.

Table~\ref{tab.1} compares the expansions ${\cal C}_7$ and ${\cal C}_9$
to the exact specific heat of the free spinless fermion
model, showing the percentage differences of the expansions of
this thermodynamic function to the exact result for
$t=1$,  $V=0$  and $\mu =0$. Table~\ref{tab.2} compares
the exact specific heat of the spin-$1/2$ Ising model,
in the presence of a longitudinal magnetic
field, in $D=1$, mapped onto the fermionic spinless Hubbard
model to the expansions ${\cal C}_7$ and ${\cal C}_9$ of
this  model, for $t=0$, $V = 0.5$ and
$\mu = 0.8$. From data in tables~\ref{tab.1} and \ref{tab.2}
we conclude that the addition of two more orders in $\beta$ in
the previous expansion of the specific heat increases the interval
in $\beta$ where this expansion is a good approximation
of the exact expression of the specific heat. Certainly, this
improvement depends on the values of the set ($t, V, \mu$).


\begin{table}[!t]
\centering 
\caption{Comparison of the percentage difference (\ref{14}) of
the expansions ${\cal C}_7$ and ${\cal C}_9$ of the specific
heat of the free spinless fermion model  for  $t=1$, $V=0$
and $\mu=0$.
}     \label{tab.1}
\vspace{2ex}
\begin{tabular}{|c| c| c|} 
\hline
\rule{0em}{3ex} $|t| \beta$ & 0.5  &  0.82 \\
\hline
\hline
\rule{0em}{3ex} $\delta_D \; {\cal C}_7 (\%)$ & -- 0.35 & -- 7.49 \\
\hline
\rule{0em}{3ex} $\delta_D\; {\cal C}_9 (\%)$ & 0.04 & 2.38  \\
\hline 
\end{tabular}
\end{table}

	
\begin{table}[!h]
\centering 
\caption{Comparison of the percentage difference (\ref{14}) of
the expansions ${\cal C}_7$ and ${\cal C}_9$ of the specific
heat corresponding to the mapping onto the spin-$1/2$ Ising
model in the presence of a longitudinal magnetic
field for $t=0$, $V= 0.5$  and $\mu=0.8$.}
\label{tab.2}
\vspace{2ex}
\begin{tabular}{|c| c| c|} 
\hline
\rule{0em}{3ex} $ |t| \beta$ & 1.6  &  1.91 \\
\hline
\hline
\rule{0em}{3ex}  $\delta_D \; {\cal C}_7 (\%)$ & -- 2.10 & -- 6.70 \\
\hline
\rule{0em}{3ex}  $\delta_D \; {\cal C}_9 (\%)$ & 0.54 & 2.32  \\
\hline 
\end{tabular}
\end{table}

Let $\rho_6 (t, V, \mu; \beta)$  and $\rho_8 (t, V, \mu; \beta)$
be the $\beta$-expansions up to order $\beta^6$ and $\beta^8$, respectively,
of the average number of spinless fermions
derived from the HFE of reference \cite{EPJB2005} and the equation (\ref{A.1}).
The effect on the convergence  $\beta$-intervals due to the terms
$\beta^7$ and $\beta^8$ in $\rho_8 (t, V, \mu; \beta)$ can
be determined by comparison of the expansions
$\rho_6 (t, V, \mu; \beta)$ and  $\rho_8 (t, V, \mu; \beta)$ to
the exact  expression of this termodynamic
function  on the mapping of the fermionic spinless Hubbard model onto on
the spin-$1/2$ Ising model, in the presence of a
longitudinal magnetic field.  In analogy to (\ref{14}), the percentage
difference regarding the functions
$\rho_6 (t, V, \mu; \beta)$, $\rho_8 (t, V, \mu; \beta)$  and
$\rho_\textrm{Ising}$ can be defined as
\begin{eqnarray}  \label{15}
\delta_D \rho_k \equiv 100\% \times
   \left( \frac{\rho_\textrm{Ising} - \rho_k}{\rho_\textrm{Ising}} \right)  ,
      \hspace{1cm}   k =7 \;\; \mbox{or} \;\; 9.
\end{eqnarray}
Here, $\rho_\textrm{Ising}$ is the mean value
of the number of spinless fermions derived from the exactly
soluble spin-$1/2$ Ising model.


\begin{table}[!b]
\centering 
\caption{Comparison of the percentage difference (\ref{15}) of the
expansions $\rho_6$ and $\rho_8$ of the of the mean
number of spinless fermions per site corresponding to the
mapping of the fermionic spinless model onto the spin-$1/2$
Ising model in the presence of a longitudinal
magnetic field for  $t= 0$, $V= 0.5$ and $\mu=0.8$.} \label{tab.3}
\vspace{2ex}
\begin{tabular}{|c| c| c|c|} 
\hline
\rule{0em}{3ex} $\beta$ & 2.5  &  2.7 & 3 \\
\hline
\hline
\rule{0em}{3ex}  $\delta_D \; \rho_6 (\%)$ & -- 0.46 & -- 0.78 & --1.62 \\
\hline
\rule{0em}{3ex}  $\delta_D \; \rho_8 (\%)$ & 0.35 & 0.67 & 1.64  \\
\hline 
\end{tabular}
\end{table}

Table~\ref{tab.3} has been generated with the percentage
difference of the expansions
$\rho_6 (t, V, \mu; \beta)$, $\rho_8 (t, V, \mu; \beta)$
to the $\rho_\textrm{Ising}$, for $t=0$, $V= 0.5$ and $\mu= 0.8$.
Data shows  that for the function $\rho (t, V, \mu; \beta)$
the presence of two orders in its $\beta$-expansion does not
really  increase the region where  the expansion is a
good approximation of the exact result, although
$\rho_8 (t, V, \mu; \beta)$ is closer to the correct result.

In general, the $\beta$-expansions of the thermodynamic functions
associated to a given model get worse as the parameters
of the Hamiltonian increase. Let us choose two sets of values
of parameters in the Hamiltonian (\ref{1a})--(\ref{1b}) that
map onto the spin-$1/2$ Ising model in the presence of a
longitudinal magnetic field:
\begin{subequations}
\begin{eqnarray}
(t= 0, V = 0.5, \mu = 0.7) &\equiv&  (1),   \label{16a}\\
(t = 0, V = 0.5, \mu = 0)  &\equiv& (2)  .   \label{16b}
\end{eqnarray}
\end{subequations}

In what follows we use the notations:
\begin{subequations}
\begin{eqnarray}
{\cal C}_9^{(1)} &\equiv& {\cal C}_9 (t= 0, V = 0.5, \mu = 0.7; \beta) ,
                   \label{17a} \\
{\cal C}_9^{(2)} &\equiv& {\cal C}_9  (t = 0, V = 0.5, \mu = 0; \beta) ,
                    \label{17b}  \\
\rho_9^{(1)} &\equiv& \rho_9 (t= 0, V = 0.5, \mu = 0.7; \beta) ,
                    \label{17c}   \\
\rho_9^{(2)} &\equiv& \rho_9  (t = 0, V = 0.5, \mu = 0; \beta)  .
                    \label{17d}
\end{eqnarray}
\end{subequations}

\begin{figure}[!h]
\begin{center}
\includegraphics[width=6.5cm,height= 6.5cm,angle= 0]{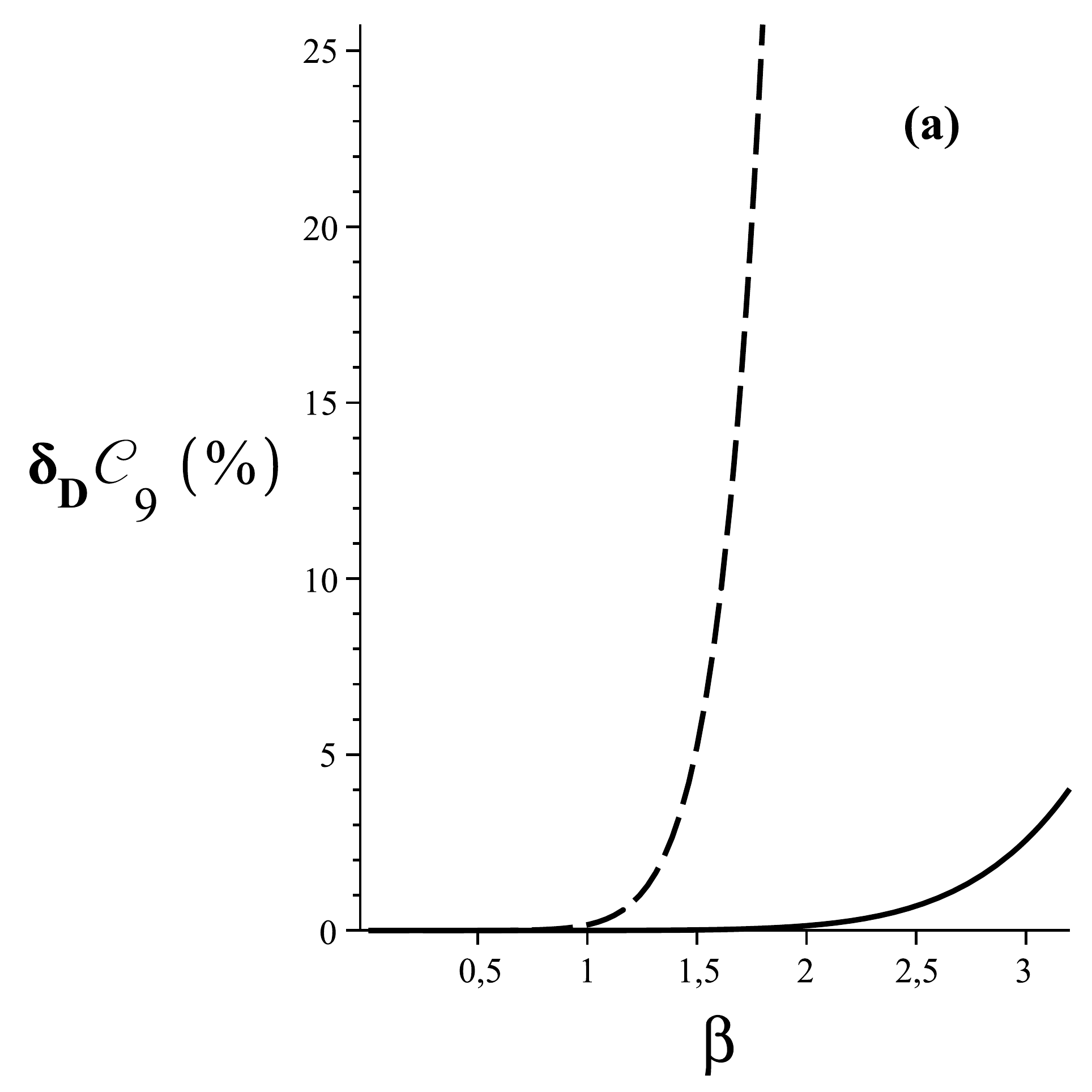}
\hspace{0.5cm}%
\includegraphics[width= 6.5cm,height= 6.5cm,angle= 0]{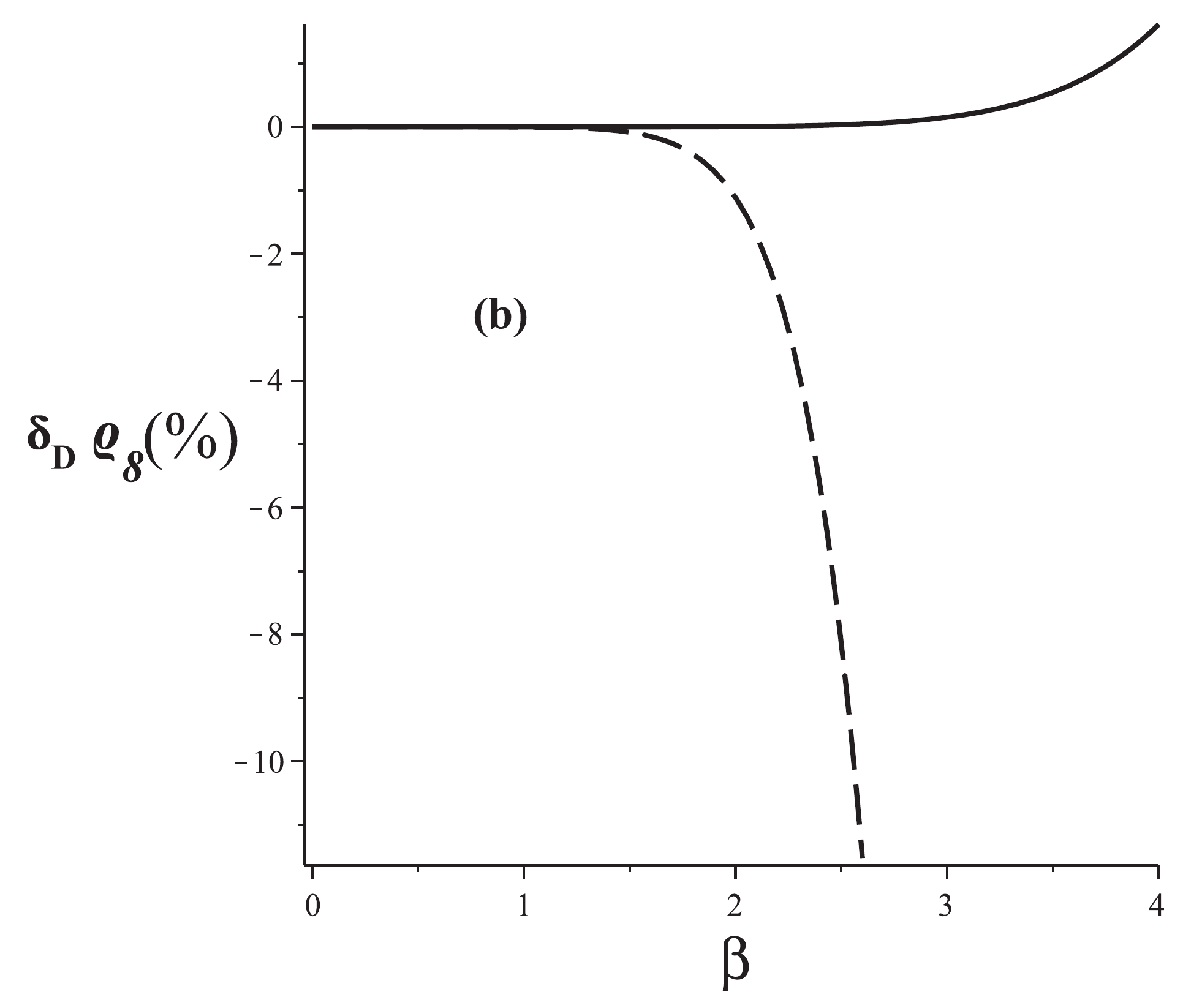}
\caption{ (a): percentage differences of ${\cal C}_9^{(1)}$ (solid line) and
${\cal C}_9^{(2)}$ (dashed line) to the  specific heat of the spin-$1/2$
Ising model. (b): percentage differences of $\rho_9^{(1)}$ (solid line) and
$\rho_9^{(2)}$ (dashed line) to the mean value of spinless fermions
also derived from the spin-$1/2$ Ising model.}
            \label{fig_1}
\end{center}
\end{figure}

Figure~\ref{fig_1} show the percentage differences
of ${\cal C}_9$ and $\rho_9$, given by (\ref{14}) and
(\ref{15}), respectively, to their respective exact expressions
for the set of values (\ref{16a}) and (\ref{16b}).
Figure~\ref{fig_1} show that for both thermodynamic
functions the percentage differences increase more rapidly for
$\mu = 0$  than for $\mu = 0.7$. How to explain
that a higher value of $\mu$ yields a larger interval in
$\beta$ where the expansions of the thermodynamic functions are
better approximations of the exact functions?

In order to understand the convergence behavior  of the expansions
of the functions ${\cal C} (t, V, \mu; \beta)$ and
$\rho (t, V, \mu; \beta)$, we define the parameter
\begin{eqnarray}  \label{18}
  x \equiv \mu - V  = h  .
\end{eqnarray}
as a measure  of how much the chain is off the
half-filling regime (i.e., $\mu = V$). Rewriting the relation (\ref{7})
between the HFE's of the one-dimensional fermionic spinless
Hubbard model and the spin-$1/2$ $XXZ$ Heisenberg model in
$D=1$ in the presence of a longitudinal magnetic field in terms of the
parameter  $x$, we obtain
\begin{eqnarray}   \label{19}
W (t, V, \mu = V + x; \beta) = - \left( \frac{V}{4} + \frac{x}{2}  \right)
+ W_{S = 1/2}\left(J = 2t, \Delta = {V}/{2t}, h = x; \beta\right)   .
\end{eqnarray}

The function $W_{S = 1/2}(J, \Delta, h; \beta)$ has a Taylor
expansion in $\beta$ whose coefficient of the $\beta^n$ term
is a product of powers of the parameters in
Hamiltonian (\ref{2}),  $J^{n_1} \Delta^{n_2} h^{n_3}$, with
$n_1 + n_2 + n_3 = n+1$. This thermodynamic function can be
written as an expansion in any of the parameters: $J, \Delta, h$ and
$\beta$. The expansion of $W_{S = 1/2}$ around $h=0=x$
corresponds to an expansion of $W (t, V, \mu = V + x; \beta)$
about the half-filling configuration, $\mu =V$.

Expanding the HFE $W (t, V, \mu = V + x; \beta)$
about $x=0$ yields
\begin{subequations}
\begin{eqnarray}   \label{20a}
W (t, V, \mu = V + x; \beta) \equiv  W (t, V, \mu = V; \beta)
     + \widetilde{W} (t, V, x; \beta)  ,
\end{eqnarray}
in which
\begin{eqnarray}
\widetilde{W} (t, V, x=0; \beta) = 0  .
\end{eqnarray}
\end{subequations}
The  symmetry relation (\ref{8}) and the definition (\ref{20a})
permit to conclude that
\begin{eqnarray}   \label{21}
\widetilde{W} (t, V, x; \beta) = - \frac{x}{2}
+ \omega (t, V, x^2; \beta)  .
\end{eqnarray}

From the form the HFE in equation (\ref{20a}) is written, one can affirm
that the thermodynamic quantities of the chain
off the half-filling regime can be
expressed as a contribution of the half-filling
configuration plus an amount due to how off the system is
from the half-filling regime (that depends, naturally, on the
parameter $x$).

The decomposition (\ref{20a}) and the definitions of the
specific heat and the mean number of fermions permit us
to write those functions in terms of the parameter $x$,
\begin{subequations}
\begin{eqnarray}   \label{22a}
{\cal C} (t, V, \mu = V+ x; \beta) =  {\cal C} (t, V, \mu = V; \beta)
  + \Delta {\cal C} (t, V,  x; \beta)   ,
\end{eqnarray}
in which
\begin{eqnarray} \label{22b}
   \Delta {\cal C} (t, V,  x; \beta) \equiv - \beta^2\;
\frac{\partial^2[ \beta \widetilde{W} (t, V, x; \beta)]}{\partial \beta^2}  ,
\end{eqnarray}
\end{subequations}
and
\begin{subequations}
\begin{eqnarray}   \label{23a}
\rho (t, V, \mu = V +x; \beta) = \frac{1}{2}
+ \Delta \rho (t, V, x; \beta)
\end{eqnarray}
with
\begin{eqnarray}
  \Delta \rho (t, V, x; \beta) \equiv - \frac{\partial}{\partial x}
      \left[  \widetilde{W} (t, V, x; \beta) + \frac{x}{2} \right]  .
\end{eqnarray}
\end{subequations}

Returning to the set of values (\ref{16a}) and (\ref{16b}) for
the parameters of Hamiltonian (\ref{1a})--(\ref{1b}) we notice
that the values of $x$ for those sets are, respectively,
\begin{eqnarray}  \label{24}
 x^{(1)} = 0.2
     \qquad \text{and} \qquad
 x^{(2)} = - 0.5  .
\end{eqnarray}
Notice that the absolute value of $x^{(1)}$ is smaller
than the absolute value of $x^{(2)}$, and this explains  why the
good approximations of those two functions are obtained in intervals
of $\beta$ that are larger for the set (\ref{16a}) than those for
the  set (\ref{16b}). This result is clearly shown in  figure~\ref{fig_1}.

In order to verify that $x$ is one of the possible parameters
of an expansion of the function $\rho(T, V, \mu= V+ x; \beta)$
rather than the chemical potential $\mu$, we calculate the
percentage weight of the $\beta^8$ term in its $\beta$-expansion.
Let us denote the $\beta$-expansion of $\Delta \rho$ by
\begin{eqnarray} \label{25}
\Delta \rho_8 (t, V, x; \beta) \equiv \sum_{l=1}^8 \;
         a_l (t, V, x)  \;  \beta^l  .
\end{eqnarray}
The percentage weight $\delta_W a_8$ of the $\beta^8$
term in the expansion of $\Delta \rho$ is
\begin{eqnarray}  \label{26}
\delta_W a_8 (t, V, x; \beta) \equiv 100\% \;
    \frac{a_8 (t, V, x)  \beta^8}{\Delta \rho (t, V, x; \beta)}  .
\end{eqnarray}


\begin{table}[!t]
\centering 
\caption{The values of $|t| \beta_\textrm{max}$ calculated from the
percentage weight $\delta a_8$ (\ref{26}) for $t=1$ and
$ {V}/{|t|} = 0.5$.}
\label{tab.4}
\vspace{2ex}
\begin{tabular}{|c|c|c|c|} 
\hline
\rule{0em}{3ex} $\frac{x}{|t|}$ & $\frac{\mu}{|t|}$ & $|t| \beta_\textrm{max}$
            & $\delta_W \; a_8(\%)$ \\
\hline
\hline
\rule{0em}{3ex} $\pm$ 0.1 & 0.4 &1.12 & -- 4.13  \\
                          & 0.6 & &   \\
\hline
\rule{0em}{3ex}  $\pm 0.5$ & 0 & 0.98 & + 4.03  \\
                             & 1 &  &   \\
\hline
\rule{0em}{3ex} $\pm 1$ & -- 0.5 &0.77 &+ 4.03  \\
                         & 1.5 &  &   \\
\hline 
\end{tabular}
\end{table}

Let $\beta_\textrm{max}$ be  the maximum value of the variable $\beta$
for which $|\delta_W a_8 (t, V, x; \beta)|\stackrel{<}{_{\sim}} 4\%$,
and for which we expect that the expansion should be still a
good approximation to the exact function $\Delta \rho(t, V, x; \beta)$.
Table~\ref{tab.4} shows the values of $|t|\beta_\textrm{max}$ and the
corresponding value of {$\delta_W a_8$} for different
values of ${x}/{|t|}$ for $t=1$ and ${V}/{|t|} = 0.5$.
The second column in this table shows the two distinct values
of $\mu$ for which the same value of $|t| \beta_\textrm{max}$ is obtained.
In particular, for ${x}/{|t|} = \pm 0.5$ we
have the chemical potentials $\mu =0$ and $\mu =1$
yielding the same value of $|t| \beta_\textrm{max}$.

In order to discuss the value of $\beta_\textrm{max}$ for which
the specific heat can be well described by its expansion,
we define the coefficients of the $\beta$-expansions of
${\cal C} (t, V, \mu= V +x;\beta)$ and of the function
$\Delta {\cal C} (t, V, x; \beta)$,
\begin{subequations}
\begin{eqnarray}  \label{27a}
{\cal C}_9 (t, V, \mu= V +x; \beta) \equiv \sum_{l=2}^9 \;\; c_l (t, V, x) \beta^l
\end{eqnarray}
and
\begin{eqnarray}   \label{27b}
\Delta {\cal C}_9 (t, V, x; \beta) \equiv \sum_{l=2}^9 \;\;
             g_l (t, V, x) \beta^l  .
\end{eqnarray}
\end{subequations}
We also define the percentage weight $\delta_W c_9$
of the term of order $\beta^9$ in the expansion ${\cal C}_9$ as
\begin{eqnarray}  \label{28}
{\delta_W c_9} (t, V, x; \beta) \equiv 100\% \;
    \frac{c_9 (t, V, x)  \beta^9}{{\cal C}_9 (t, V, V + x; \beta)}  ,
\end{eqnarray}
in order to determine the value of $\beta_\textrm{max}$ for the
specific heat.


\begin{table}[!b]
\centering 
\caption{  The percentage weight $\delta_W c_9$ of the term of
order $\beta^9$ in the expansion ${\cal C}_9 (t, V, V + x; \beta)$.
The values of  $|t| \beta_\textrm{max}$ are calculated for $t=1$ and
${V}/{|t|} = 0.5$.}
\label{tab.5}
\vspace{2ex}
\begin{tabular}{|c|c|c|c|} 
\hline
\rule{0em}{3ex} $\frac{x}{|t|}$ & $\frac{\mu}{|t|}$ & $|t| \beta_\textrm{max}$
        & $\delta_W \; c_9(\%)$ \\
\hline
\hline
\rule{0em}{3ex} $\pm$ 0.1 & 0.4 & 0.68 & -- 4.10  \\
                          & 0.6 & &   \\
\hline
\rule{0em}{3ex}  $\pm 0.5$ & 0 & 0.69 & -- 4.06  \\
                             & 1 &  &   \\
\hline
\rule{0em}{3ex} $\pm 1$ & -- 0.5 & 0.65 & + 4.05  \\
                         & 1.5 &  &   \\
\hline 
\end{tabular}
\end{table}

Table~\ref{tab.5}{} shows the values of $|t| \beta_\textrm{max}$ for
the specific heat where {$\delta_W c_9 \stackrel{<}{_{\sim}} 4\%$.}
The calculations have been done with $t=1$ and ${V}/{|t|} = 0.5$.
Again we obtain that for $x = \pm 0.5$,  the $\beta$- interval,
where the expansion  ${\cal C}_9$ is a good approximation
of the exact expression  of this thermodynamic function, is the same
for $\mu = 0$ and $\mu = 1$.

The function $\Delta {\cal C}$ in equation (\ref{22b}) measures the difference
between the specific heat in the half-filling regime  and
that function at the chemical potential $\mu = V+ x$. It also
has a $\beta$-expansion that depends on $x$. In order to verify the
value of $\beta_\textrm{max}$ for the function  $\Delta {\cal C}_9$,
we define the percentage weight of the term of order
$\beta^9$ in this function,
\begin{eqnarray}  \label{29}
\delta_W g_9 (t, V, x; \beta) \equiv 100\% \;
    \frac{g_9 (t, V, x)  \beta^9}{\Delta {\cal C}_9 (t, V, x; \beta)}  .
\end{eqnarray}
Table~\ref{tab.6} shows
the values of $|t| \beta_\textrm{max}$  for the function
$\Delta {\cal C}_9 (t, V, x; \beta)$ with $|t| =1$ and
${V}/{|t|} = 0.5$. We verify that the values
of $|t| \beta_\textrm{max}$ for the functions ${\cal C}_9 (t, V, V + x; \beta)$ and
$\Delta {\cal C}_9 (t, V, x;\beta)$  can be different.

\begin{table}[!t]
\centering 
\caption{The percentage weight $\delta_W g_9 (\%)$ of
the term of order $\beta^9$ in  the expansion of
$\Delta {\cal C}_9 (t, V, V + x; \beta)$.
The values of $|t| \beta_\textrm{max}$ are calculated
for $t=1$ and ${V}/{|t|} =  0.5$.
}
\label{tab.6}
\vspace{2ex}
\begin{tabular}{|c|c|c|c|} 
\hline
\rule{0em}{3ex} $\frac{x}{|t|}$ & $\frac{\mu}{|t|}$ & $|t| \beta_\textrm{max}$
     & $\delta_W g_9 (\%)$ \\
\hline
\hline
\rule{0em}{3ex} $\pm$ 0.1 & 0.4 & 0.56 & -- 4.03  \\
                          & 0.6 & &   \\
\hline
\rule{0em}{3ex}  $\pm 0.5$ & 0 & 0.56 & + 4.05  \\
                             & 1 &  &   \\
\hline
\rule{0em}{3ex} $\pm 1$ & -- 0.5 & 0.47 & + 4.00  \\
                         & 1.5 &  &   \\
\hline 
\end{tabular}
\end{table}


\section{The temperature dependence of the chemical potential}  \label{S4}

The chemical potential $\mu$ is one of the parameters
in the Hamiltonian  (\ref{1a})--(\ref{1b}). For a given fixed
value of $\mu$, the relation
$\rho (t, V, \mu; \beta) = - {\partial W(t, V, \mu; \beta)}/{\partial \mu}$
permits the determination, from  the expansion (\ref{A.1}), how
the mean number of  spinless fermions varies with the temperature.

\begin{figure}[!b]
\begin{center}
\includegraphics[width= 6.cm,angle= 0]{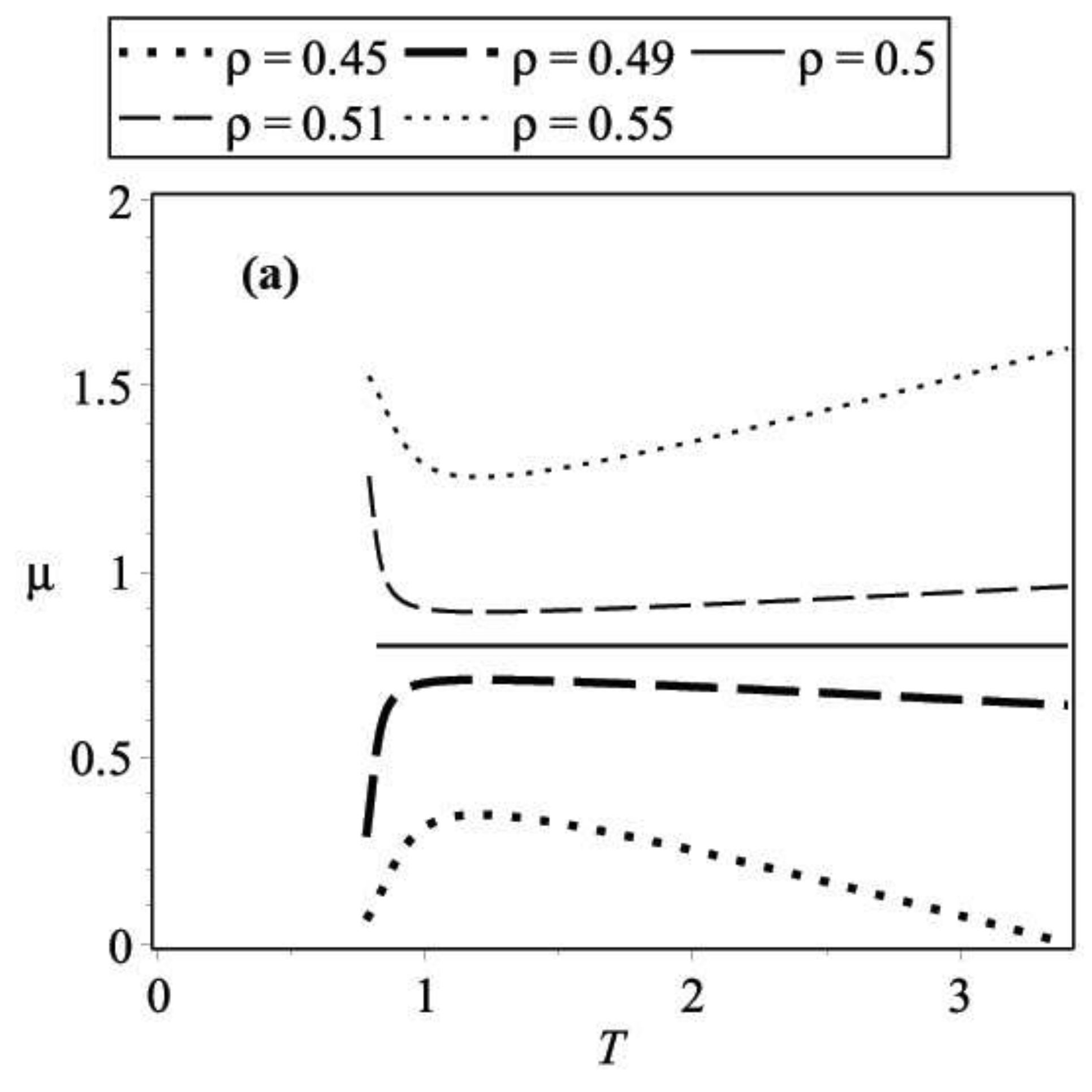}
\hspace{0.5cm}
\includegraphics[width=6.cm,angle= 0]{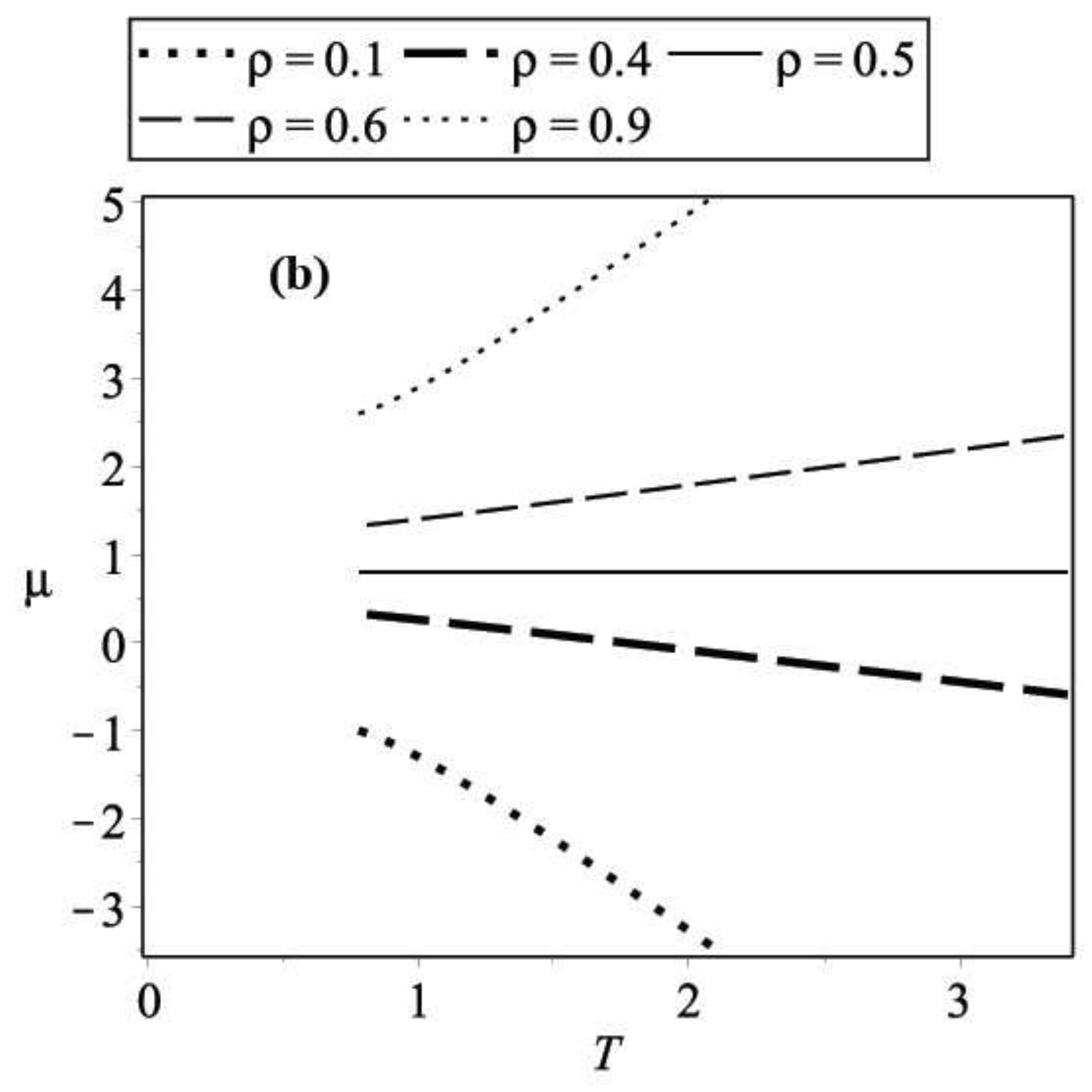}
\caption{The chemical potential $\mu (t, V, \rho_8; T)$ as
function of the temperature $T$. (a): for $t=1$ and
${V}/{|t|} = 0.8$. (b): for $t=0$ and
${V}/{|t|} = 1$.}
	    \label{fig_2}
\end{center}
\end{figure}

How should the chemical potential vary for a given temperature $T$,
keeping the chain in thermal equilibrium at this temperature,
so that the chain keeps its number of fermions per site?
The relation between $\rho (t, V, \mu; \beta)$ and $W(t, V, \mu; \beta)$
permits to rewrite the expansion $\rho_8 (t, V, \mu; \beta)$  as
a polynomial in the chemical potential $\mu$ of order $\mu^7$,
written as
\begin{eqnarray}   \label{30}
\rho_8 = n_0 (t, V; \beta) \mu^0 + n_1 (t, V; \beta) \mu^1 + \cdots
+ n_7 (t, V; \beta) \mu^7  .
\end{eqnarray}
The coefficients  $n_l (t, V; \beta)$, with
$l \in\{ 0, 1, \dots, 7\}$,
are known and~--- differently from the coefficients of the $\beta$-terms
in the expansion (\ref{A.1})~--- they have corrections
from higher orders in $\beta$.

In order to derive the dependence of the function $\mu$ on the variables
$\rho_8$, $t, V$ and $\beta$, one must obtain the roots of a
$7^{\rm th}$ degree polynomial in $\mu$.
Figure~\ref{fig_2} show  our numerical results for the dependence
of $\mu$ on the temperature  $T$ for $t=0$ and $t=1$,  for fixed values
of ${V}/{|t|}$ and $\rho_8$.

By comparing the curves in each graph of figure~\ref{fig_2},
we obtain the relation
\begin{eqnarray}  \label{31}
\mu (t, V, \rho = 0.5 + \delta; T) = 2V - \mu (t, V, \rho = 0.5 - \delta; \beta)  ,
\end{eqnarray}
with $\delta \in [ -0.5, 0.5]$. This is similar to
equation (\ref{9}), derived from  the hole-particle symmetry of the
one-dimensional fermionic spinless Hubbard model.

\begin{figure}[!t]
\begin{center}
\includegraphics[width= 5.5cm,height= 6.5cm,angle= 0]{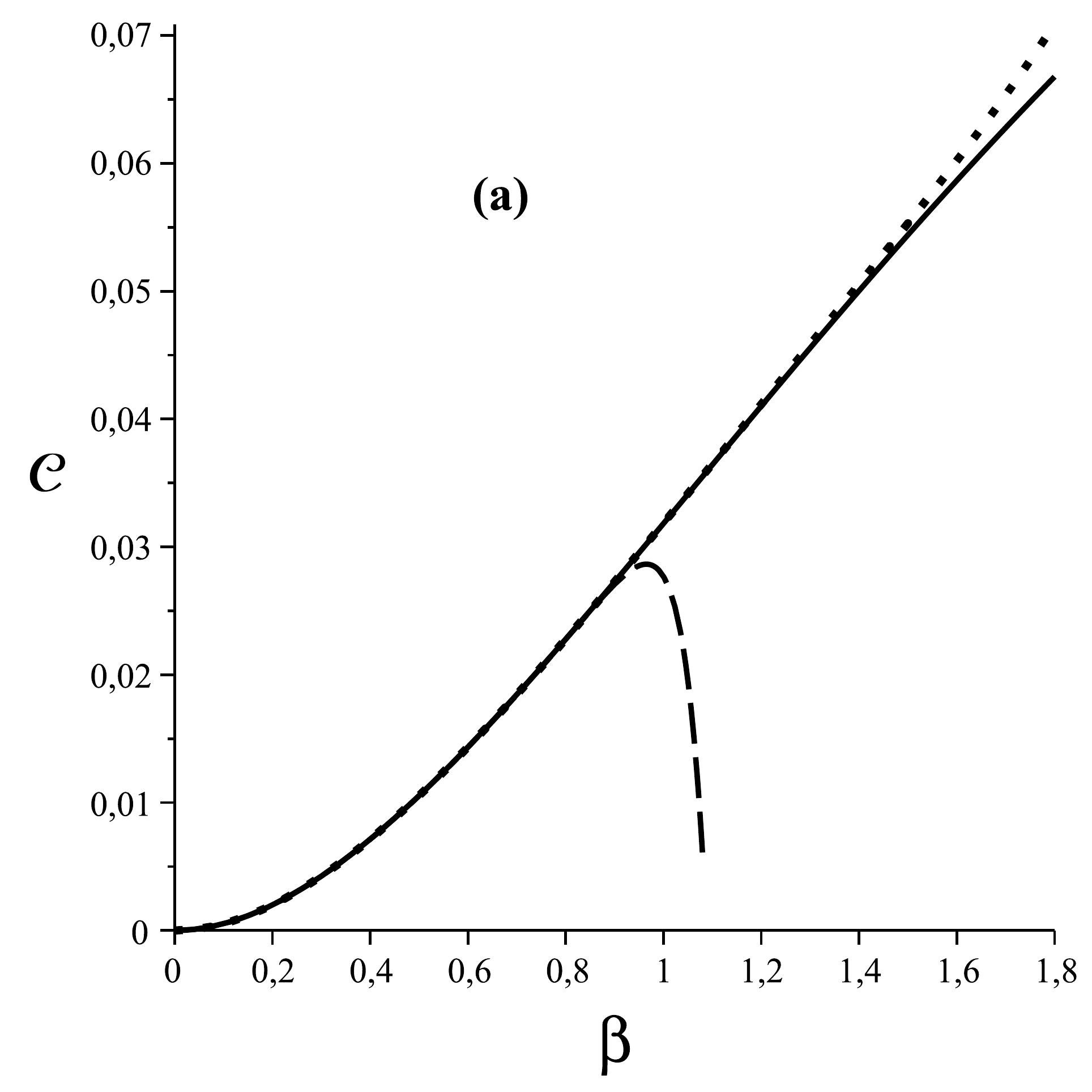}
\hspace{0.5cm}
\includegraphics[width= 5.5cm,height= 6.5cm,angle= 0]{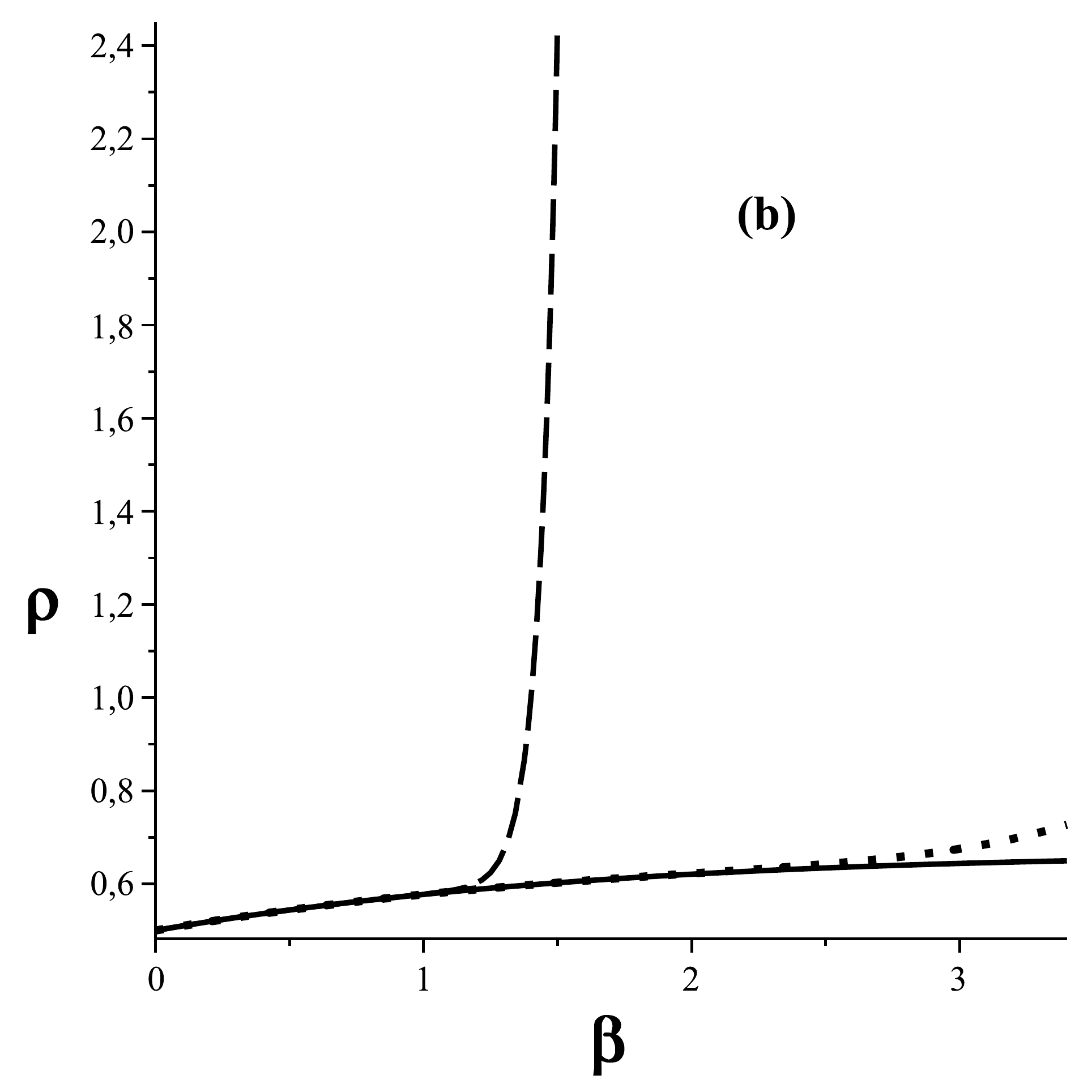}
\caption{(a): comparison of the specific heat curves ${\cal C}_\textrm{Ising}$ (solid line),
${\cal C}_8$ (dotted line) and ${\cal C}_{24}$ (dashed line). (b): comparison of the
average number of fermions $\rho_\textrm{Ising}$ (solid line),
$\rho_7$ (dotted line) and $\rho_{23}$ (dashed line). In both
panels, $t=0$, $V = 0.5$ and $\mu = 0.9$.
 }
 	    \label{fig_3}
\end{center}
\end{figure}


\section{Conclusions}   \label{S6}

The one-dimensional fermionic spinless fermionic Hubbard model is the simplest
fermionic  model, and it has the particle-hole symmetry. This model can be
mapped onto the spin-$1/2$ $XXZ$ Heisenberg model in the presence of
a longitudinal magnetic field in $D=1$. Some years ago  we derived the
$\beta$-expansion of the HFE of the latter up to order $\beta^6$ \cite{EPJB2005}. In
this article we have extended  the $\beta$-expansion of the HFE of both models
up to order $\beta^8$. Each $\beta$ term in the expansion
satisfies the condition (\ref{10})  derived from the particle-hole
symmetry of the one-dimensional fermionic model.

We have used the expansion (\ref{A.1}) of the HFE of the
fermionic  spinless model (\ref{1a})--(\ref{1b}) to study how
the interval of convergence (in $\beta$) of  the specific heat
per site [${\cal C} (t, V, \mu; \beta)$] and
of the mean number of spinless fermions per site
[$\rho (t, V, \mu; \beta)$] is modified by the
presence of two more orders in $\beta$ in their respective
expansions.

An interesting result that we obtain for the $\beta$ expansions
of the thermodynamic functions comes from the relation (\ref{7})
between the HFE of the fermionic spinless model and the spin-$1/2$
model. When the chain is off the half-filling regime
($\mu \not= V$), the relation (\ref{7}) permits to write the thermodynamic
functions of the chain in this regime as two $\beta$-expansions:
the expansion of the function in the half-filling ($\mu =V$)
plus another expansion that depends on the
set of parameters ($t, V, x= \mu - V; \beta)$.
The parameter $x$ is a measure of how off the chain is from the
half-filling regime. This fact explains why the expansions
${\cal C}_9 (t, V, \mu; \beta)$ and $\rho_8 (t, V, \mu; \beta)$
with $\mu =0$  have shorter $\beta$ intervals of convergence
than those for $|\mu| > 0$.

We have numerically obtained  the dependence of the chemical
potential $\mu$ on the temperature $T$ when the mean value of
fermions per site $\rho$ is kept fixed. We have verified that
the relation (\ref{31}),  satisfied by $\mu (T)$
for $\rho = 0.5 \pm \delta$ with  $\delta \in [ -0.5, +0.5]$,
is similar to equation (\ref{9}) derived from  the particle-hole symmetry
of the fermionic spinless  Hubbard model  in $D=1$.

Finally, we point out that the present $\beta$-expansion
of the HFE of the one-dimensional spinless Hubbard model is valid
for any set of parameters of its Hamiltonian,  including
the cases $V>0$ (repulsion) and $V<0$ (attraction).

\section*{Acknowledgements}

O. Rojas thanks CNPq and FAPEMIG for partial financial support.

\appendix

\setcounter{equation}{0}
\def\theequation{A.\arabic{equation}}

\section{The HFE of the one-dimensional fermionic spinless  Hubbard
model up to order $\beta^8$}  \label{Apend_A}

We have applied the method of reference \cite{chain_m} to calculate
the $\beta$-expansion of the HFE associated to the
Hamiltonian (\ref{1a})--(\ref{1b}) and  to the Hamiltonian (\ref{2})
[see equation (\ref{7})]. We  have also implemented a new set of rules
that permit the algebraic computation  of the HFE of the  one-dimensional
fermionic spinless  Hubbard model up to order  $\beta^8$,
%
\begin{eqnarray}
  W(t, V, \mu; \beta) = \!\! &-& \frac {\ln  \left( 2 \right) }{\beta}
+ \frac{1}{4}\,V - \frac{1}{2}\,\mu
+ \left( -\frac {5}{32}\,{V}^{2}+1/4\,V\mu
 \frac{1}{8}\,{\mu}^{2}- \frac{1}{4}\,{t}^{2} \right) \beta
              \nonumber \\
&+& \left( \frac{1}{16}\,{V}^{3}
- \frac{1}{8}\,{V}^{2}\mu
+ \frac{1}{16}\,V{\mu}^{2}
- \frac{1}{16}\,V{t}^{2} \right) {\beta}^{2}
+ \left( - \frac{31}{3072} \,{V}^{4}
+ \frac {1}{96} \,{V}^{3}\mu
+ \frac {1}{64} \,{V}^{2}{\mu}^{2}   \right. \nonumber \\
&+&  \left.
 \frac {7}{96}\,{V}^{2}{t}^{2}
- \frac{1}{48}\,V{\mu}^{3}
-1/8\,V\mu\,{t}^{2}
+ \frac {1}{192} \,{\mu}^{4}
+ \frac{1}{16}\,{\mu}^{2}{t}^{2}
+ \frac{1}{32}\,{t}^{4} \right) {\beta}^{3}
+ \left( -{\frac {1}{128}}\,{V}^{5}    \right.   \nonumber   \\
&+& {\frac {7}{192}}\,{V}^{4}\mu-{\frac {23}{384}}\,{V}^{3}{\mu}^{2}
- {\frac {7}{256}}\,{V}^{3}{t}^{2}+\frac{1}{24}\,{V}^{2}{\mu}^{3}
+ \frac{1}{16}\,{V}^{2}\mu\,{t}^{2}-{\frac {1}{96}}\,V{\mu}^{4}
- \frac{1}{32}\,V{\mu}^{2}{t}^{2}   \nonumber \\
&+& \left.  \frac{1}{32}\,V{t}^{4} \right) {\beta}^{4}
+ \left( {\frac {287}{36864}}\,{V}^{6}
-{\frac {239}{7680}}\,{V}^{5}\mu
+{\frac {139}{3072}}\,{V}^{4}{\mu}^{2}    \right.
- {\frac {21}{2560}}\,{V}^{4}{t}^{2}
-{\frac {31}{1152}}\,{V}^{3}{\mu}^{3}   \nonumber  \\
&+& {\frac {7}{192}}\,{V}^{3}\mu\,{t}^{2}
+  {\frac {5}{1536}}\,{V}^{2}{\mu}^{4}
-{\frac {23}{384}}\,{V}^{2}{\mu}^{2}{t}^{2}
-{\frac {47}{1536}}\,{V}^{2}{t}^{4}
+{\frac {1}{480}}\,V{\mu}^{5}
+ \frac{1}{24}\,V{\mu}^{3}{t}^{2}      \nonumber  \\
&+& \frac{1}{16}\,V\mu\,{t}^{4}
-{\frac {1}{2880}}\,{\mu}^{6}
 \left.
- {\frac {1}{96}}\,{\mu}^{4}{t}^{2}
-\frac{1}{32}\,{\mu}^{2}{t}^{4}
-{\frac {1}{144}}\,{t}^{6} \right) {\beta}^{5}
+ \left( -{\frac {29}{10240}}\,{V}^{7}  \right.    \nonumber   \\
&+& {\frac {389}{46080}}\,{V}^{6}\mu
- {\frac {119}{30720}}\,{V}^{5}{\mu}^{2}
+ {\frac {1603}{92160}}\,{V}^{5}{t}^{2}
-{\frac {7}{576}}\,{V}^{4}{\mu}^{3}
-{\frac {53}{768}}\,{V}^{4}\mu\,{t}^{2}
+ {\frac {41}{2304}}\,{V}^{3}{\mu}^{4}     \nonumber \\
&+& {\frac {157}{1536}}\,{V}^{3}{\mu}^{2}{t}^{2}
+{\frac {83}{23040}}\,{V}^{3}{t}^{4}
- {\frac {17}{1920}}\,{V}^{2}{\mu}^{5}
-{\frac {13}{192}}\,{V}^{2}{\mu}^{3}{t}^{2}
-{\frac {1}{64}}\,{V}^{2}\mu\,{t}^{4}
+ {\frac {17}{11520}}\,V{\mu}^{6}      \nonumber \\
&+& {\frac {13}{768}}\,V{\mu}^{4}{t}^{2}
+{\frac {1}{128}}\,V{\mu}^{2}{t}^{4}
 \left.
- {\frac {11}{768}}\,V{t}^{6} \right) {\beta}^{6}
+ \left( -{\frac {108527}{165150720}}\,{V}^{8}
+{\frac {17}{645120}}\,{\mu}^{8}     \right.  \nonumber \\
&+&  {\frac {17}{9216}}\,{t}^{8}
+ {\frac {1709}{46080}}\,{V}^{5}\mu\,{t}^{2}
-{\frac {1439}{30720}}\,{V}^{4}{\mu}^{2}{t}^{2}
+{\frac {19}{1152}}\,{V}^{3}{\mu}^{3}{t}^{2}
-{\frac {83}{1536}}\,{V}^{3}\mu\,{t}^{4}  \nonumber   \\
&+& {\frac {49}{4608}}\,{V}^{2}{\mu}^{4}{t}^{2}
+  {\frac {73}{1024}}\,{V}^{2}{\mu}^{2}{t}^{4}
-{\frac {17}{1920}}\,V{\mu}^{5}{t}^{2}
-{\frac {17}{384}}\,V{\mu}^{3}{t}^{4}
- {\frac {17}{576}}\,V\mu\,{t}^{6}     \nonumber\\
&+&{\frac {9241}{1290240}}\,{V}^{7}\mu
-{\frac {8939}{368640}}\,{V}^{6}{\mu}^{2}
- {\frac {12877}{1290240}}\,{V}^{6}{t}^{2}
+{\frac {3539}{92160}}\,{V}^{5}{\mu}^{3}
- {\frac {2311}{73728}}\,{V}^{4}{\mu}^{4}  \nonumber   \\
&+& {\frac {191}{12288}}\,{V}^{4}{t}^{4}
+ {\frac {287}{23040}}\,{V}^{3}{\mu}^{5}
-{\frac {73}{46080}}\,{V}^{2}{\mu}^{6}
+{\frac {31}{2880}}\,{V}^{2}{t}^{6}
- {\frac {17}{80640}}\,V{\mu}^{7}  \nonumber  \\
&+& {\frac {17}{11520}}\,{\mu}^{6}{t}^{2}
+{\frac {17}{1536}}\,{\mu}^{4}{t}^{4}
\left.+ {\frac {17}{1152}}\,{\mu}^{2}{t}^{6} \right) {\beta}^{7}
+ \left( {\frac {817}{92160}}\,{V}^{6}\mu\,{t}^{2}
-{\frac {841}{20480}}\,{V}^{5}{\mu}^{2}{t}^{2}    \right.  \nonumber \\
&+& {\frac {85}{1152}}\,{V}^{4}{\mu}^{3}{t}^{2}
+{\frac {473}{7680}}\,{V}^{4}\mu\,{t}^{4}
-{\frac {299}{4608}}\,{V}^{3}{\mu}^{4}{t}^{2}
- {\frac {1393}{15360}}\,{V}^{3}{\mu}^{2}{t}^{4}
+{\frac {107}{3840}}\,{V}^{2}{\mu}^{5}{t}^{2}  \nonumber\\
%
&+&  {\frac {23}{384}}\,{V}^{2}{\mu}^{3}{t}^{4}
-{\frac {5}{1152}}\,{V}^{2}\mu\,{t}^{6}
-{\frac {107}{23040}}\,V{\mu}^{6}{t}^{2}
- {\frac {23}{1536}}\,V{\mu}^{4}{t}^{4}
+{\frac {5}{2304}}\,V{\mu}^{2}{t}^{6}    \nonumber   \\
&+& {\frac {4619}{3096576}}\,{V}^{9}
- {\frac {24649}{2580480}}\,{V}^{8}\mu
+{\frac {125497}{5160960}}\,{V}^{7}{\mu}^{2}
+ {\frac {367}{4128768}}\,{V}^{7}{t}^{2}
-{\frac {4259}{138240}}\,{V}^{6}{\mu}^{3}    \nonumber  \\
&+& {\frac {3449}{184320}}\,{V}^{5}{\mu}^{4}
-{\frac {9991}{645120}}\,{V}^{5}{t}^{4}
- {\frac {53}{23040}}\,{V}^{4}{\mu}^{5}
-{\frac {443}{138240}}\,{V}^{3}{\mu}^{6}
+{\frac {65}{13824}}\,{V}^{3}{t}^{6}   \nonumber   \\
	&+& \left. {\frac {31}{20160}}\,{V}^{2}{\mu}^{7}
	-{\frac {31}{161280}}\,V{\mu}^{8}
	+{\frac {29}{4608}}\,V{t}^{8}
	  \right) {\beta}^{8}  + {\cal O} (\beta^9).
\label{A.1}
\end{eqnarray}


\ukrainianpart

\title{$\beta$-розвинення  $D=1$ ферміонної безспінової моделі Габбарда поза половинним заповненням}

\author{Е.В. Кореа Сільва\refaddr{lb1}, М.Т. Томаз\refaddr{lb2}, O. Рохас\refaddr{lb3}}

\addresses{
\addr{lb1}
Технологічний факультет, Державний університет м. Ріо-де-Жанейро, Резенді, Бразилія
\addr{lb2} Інститут фізики, Федеральний університет Флуміненсе, Нітерой, Бразилія
\addr{lb3} Відділ фізики, Федеральний університет Лаврас, Лаврас,  Бразилія
}

\vspace{-8mm}
\makeukrtitle

\begin{abstract}
Встановлено, що для безспінової моделі поза половинним заповненням ($\mu\neq
 V$) вільну енергію Гельмгольца можна записати у вигляді двох
 $\beta$-розвинень: одне розвинення походить від конфігурації з половинним
 заповнення, а інше залежить від параметра відхилення $x = \mu - V$. Чисельно
 показано, що хімічний потенціал як функція температури задовольняє
 співвідношення подібне до того, яке отримується з симетрії частинка-дірка
 ферміонної безспінової моделі. $\beta$-розвинення вільної енергії Гельмгольца
 одновимірної ферміонної безспінової моделі Габбарда продовжено аж до порядку
 $\beta^8$.
\keywords квантова статистична механіка, сильно скорельвана електронна система, моделі спінових ланцюжків
\end{abstract}
\end{document}